\documentclass[preprint,preprintnumbers,amsmath,amssymb]{revtex4-1}

\usepackage{graphicx,tabularx}
\usepackage{bm}
\usepackage{float}
\usepackage{dcolumn}
\usepackage{color}
\usepackage{ulem}
\usepackage{mathptmx}
\usepackage{moreverb}
\usepackage{hyperref}

\hypersetup{hidelinks,
       colorlinks=true,
       allcolors=blue,
       pdfstartview=Fit,
       breaklinks=true}

\renewcommand{\thefigure}{\arabic{figure}}


\begin{document}
\renewcommand{\figurename}{\textbf{Fig.}}
\renewcommand{\thefigure}{\textbf{\arabic{figure}}}

\title{
Abnormally enhanced Hall Lorenz number in the magnetic Weyl semimetal NdAlSi
}

\author{Nan Zhang$^1$, Daifeng Tu$^{1,2}$, Ding Li$^{1,2}$, Kaixin Tang$^3$, Linpeng Nie$^3$,  Houpu Li$^1$, Hongyu Li$^3$, Tao Qi$^1$, Tao Wu$^{3,4}$}
\author{Jianhui Zhou$^{2}$}
\email{jhzhou@hmfl.ac.cn}
\author{Ziji Xiang$^{3,4}$}
\email{zijixiang@ustc.edu.cn}
\author{Xianhui Chen$^{1,3,4}$}
\email{chenxh@ustc.edu.cn}

\affiliation{
$^1$Department of Physics, University of Science and Technology of China, Hefei, Anhui 230026, China\\
$^2$Anhui Key Laboratory of Low-Energy Quantum Materials and Devices, High Magnetic Field Laboratory, HFIPS, Anhui, Chinese Academy of Sciences, Hefei, Anhui 230031, China\\
$^3$Hefei National Research Center for Physical Sciences at the Microscale, University of Science and Technology of China, Hefei, Anhui 230026, China.\\
$^4$Hefei National Laboratory, University of Science and Technology of China, Hefei, Anhui 230088, China.
}
\date{\today}

\maketitle                   
\noindent
\textbf{In Landau's celebrated Fermi liquid theory, electrons in a metal obey the Wiedemann--Franz law at the lowest temperatures. This law states that electron heat and charge transport are linked by a constant $L_0$, i.e., the Sommerfeld value of the Lorenz number ($L$). Such relation can be violated at elevated temperatures where the abundant inelastic scattering leads to a reduction of the Lorenz number ($L < L_0$). Here, we report a rare case of remarkably enhanced Lorenz number ($L > L_0$) discovered in the magnetic topological semimetal NdAlSi. Measurements of the transverse electrical and thermal transport coefficients reveal that the Hall Lorenz number $L_{xy}$ in NdAlSi starts to deviate from the canonical value far above its magnetic ordering temperature. Moreover, $L_{xy}$ displays strong nonmonotonic temperature and field dependence, reaching its maximum value close to 2$L_0$ in an intermediate parameter range. Further analysis excludes charge-neutral excitations as the origin of enhanced $L_{xy}$. Alternatively, we attribute it to the Kondo-type elastic scattering off localized 4$f$ electrons, which creates a peculiar energy distribution of the quasiparticle relaxation time.  Our results provide insights into the perplexing transport phenomena caused
by the interplay between charge and spin degrees of freedom.}

\bigskip
\noindent
{\textbf{\large Introduction}}

\noindent
For the majority of metallic systems, the electronic excitations therein can be described by the concept of quasiparticles in Landau's Fermi liquid picture. At sufficiently low temperatures ($T$) where the life time of quasiparticles is predominantly limited by the elastic scattering on impurities or lattice defects, the Fermi liquid theory guarantees an elegant relationship between the thermal and electrical conductivities ($\kappa$ and $\sigma$, respectively) contributed by quasiparticles: $\kappa/\sigma T = L$. Here, the parameter $L$ is the Lorenz number which is predicted to take the Sommerfeld value $L_0$ = $(\pi^2/3)(k_B/e)^2$ = 2.44$\times$10$^{-8}$\,W $\Omega$ K$^{-2}$ ($k_B$ is the Boltzmann constant) \cite{BookZiman}. This is the famous Wiedemann--Franz (WF) law established over one and a half centuries ago. The fact that the WF law prevails in almost all known metallic compounds is indeed not surprising: the elastic impurity scattering, always dominant at the zero-temperature limit, impedes the electrical (charge) and thermal (energy) transport in exactly the same momentum relaxing process. Very rare exceptions do exist with the value of $L$ distinct from $L_0$ down to the lowest $T$; such behaviour is usually taken as evidence for the failure of quasiparticle description \cite{HartnollPRB}, highlighting the non-Fermi-liquid physics introduced by critical fluctuations \cite{TanatarCeCoIn5,KondoBreakdown} or one dimensionality \cite{Bechgaard,Li0.9Mo6O17}.

Despite the robustness of the WF law at $T$ = 0 limit, a violation constantly occurs at intermediate $T$ (ranging from a few kelvins to approximately the Debye temperature) when the inelastic scattering of quasiparticles sets in \cite{BookBerman}. The inelastic electron-phonon ($e-ph$) and electron-electron ($e-e$) scattering processes relax the charge and energy flows in different rates, thus causing a $T$-dependent Lorenz number $L(T) \neq L_0$ \cite{DasSarmaPRB,PrincipiPRL}. As revealed by recent investigations, several scattering mechanisms can lead to severe finite-$T$ deviation from the WF law, including small-angle (nearly momentum-conserving) $e-ph$ and $e-e$ scattering which may give rise to hydrodynamic transport \cite{PrincipiPRL,LucasPRB,GoothWP2,BehniaWP2,BehniaAntimony} and interband $e-e$ scattering that can be important in semimetals \cite{SongciLi,ZareniaPRB}. In spite of the diversity of microscopic mechanisms, the above-mentioned inelastic scattering effects invariably cause a reduction of $L$ in three-dimensional (3D) materials \cite{DasSarmaPRB,PrincipiPRL,ZareniaPRB}, i.e., $L < L_0$ for intermediate $T$.

In this paper, we report an unexpectedly enhanced $L$---contrary to the case in most metals and semimetals \cite{BehniaWP2,BehniaAntimony,Cu-YBCO,CeRhIn5}---in a magnetic Weyl semimetal, NdAlSi. The topologically nontrivial band structure with multiple Weyl nodes in NdAlSi has been verified by both the density-functional theory (DFT) calculations \cite{NdAlSiNeutron,NdAlSiChen,NdAlSiQO} and the angle-resolved photoemission spectroscopy (ARPES) measurements \cite{NdAlSiARPES}. The Weyl semimetal state hosting totally 40 Weyl nodes in paramagnetic NdAlSi is promised by the noncentrosymmetric space group $I$$4_1$$md$ \cite{NdAlSiARPES}, whereas the magnetic-field($H$)-induced spin-polarized state is also identified as a Weyl semimetal possessing a total of 56 Weyl nodes \cite{NdAlSiNeutron}; some of these Weyl nodes are close to the Fermi level ($E_F$), as two of of the four Fermi surface pockets host Weyl fermions and are characterized by linear dispersions \cite{NdAlSiNernst}. In NdAlSi, the Nd$^{3+}$ (4$f^3$) local moments form a helical ferrimagnetic order below $T_m$ = 7.3\,K, which is proposed to emerge from the Ruderman-Kittel-Kasuya-Yosida (RKKY) interaction between 4$f$ moments, mediated by the Weyl fermions (predominantly Nd 5$d$ electrons) \cite{NdAlSiNeutron}. The helical ferrimagnetism, characterized by a unique up-down-down (u-d-d) spin configuration \cite{NdAlSiNeutron}, can be destructed by a magnetic field ($H$) along the crystallographic $c$ ([001]) direction at a metamagnetic transition $H = H_m$ and a fully-polarized state is stabilized above a crossover field $H_p$ (Fig.\,1b) \cite{NdAlSiQO,NdAlSiChen,NdAlSiQOchen}. The complex magnetic phases enriched by their interplays with low-energy quasiparticle excitations leave unique fingerprints in thermal transport. The behaviour of longitudinal thermal conductivity $\kappa_{xx}$ implies that the phonon heat conduction is strongly affected by scattering on local spins and magnetic excitations. More importantly, by scrutinizing the thermal Hall effect $\kappa_{xy}$, we revealed an enhancement of $L$ exceeding $L_0$ which develops at $\sim$ 30\,K upon cooling, well above $T_m$; the enhanced $L$ persists over a wide range of $H$ and is only reduced in the polarized state. We proved that the unusually large $L$ is not contributed by charge-neutral excitations, e.g., phonons and magnons. Instead, the finite-$T$ deviation from WF law in NdAlSi can be interpreted by a model considering a Kondo-type scattering between localized and itinerant electrons. Such scattering mechanism, while being overlooked in most of the previous studies, may indeed play a role in a wide range of metallic compounds that possess local magnetic moments.


\bigskip
\noindent
{\textbf{\large Results}}

\noindent
\textbf{\normalsize Longitudinal thermal conductivity}

\noindent
We first look into the coupling between the spin and lattice systems. Figure\,1d shows the zero-field $\kappa_{xx} (T)$ and $\kappa_{zz} (T)$, measured with heat current $J_Q$ applied along crystallographic $a$ ([100]) and $c$ axes, respectively (see Methods for details; calibrations of the thermal-transport experimental setup are elaborated on in Supplementary Note 1). The electronic contribution $\kappa_e$, roughly estimated here by using the WF value $\kappa_{xx}^{\rm WF}= L_0\sigma_{xx} T$ (solid lines in Fig.\,1d), can account for less than one half of the total $\kappa$ for both directions. The remaining part for $T >$ 20\,K (see below) can be taken as the phonon contribution $\kappa_{ph} (T)$; it exhibits a glasslike behavior characterized by the lack of a peak profile and the weak $T$ dependence at high temperature \cite{BookBerman}. The absence of phonon peak implies excessive scattering experienced by phonons at the temperature of several tens of kelvin---much above $T_m$. In contrast, a pronounced phonon peak on $\kappa_{xx}$ occurs at $\sim$30\,K in a nonmagnetic isostructural compound LaAlSi (Fig.\,1e), highlighting the crucial role played by Nd 4$f$ moments in the suppression of $\kappa_{ph}$.

A likely cause of such suppression in the paramagnetic state of NdAlSi is the resonant phonon scattering off the crystal-electric-field (CEF) levels of the 4$f$ multiplets \cite{NdMnO3,Pr2Ir2O7}. In this scenario, the CEF multiplets absorb and emit phonons with the energy matching their level splitting, effectively impeding phonon heat conduction since the absorbed and emitted phonons have unrelated momenta. The CEF configuration for Nd$^{3+}$ at zero field is composed of a ground state doublet and four degenerate excited doublets with an energy splitting $\sim$ 4.1 meV \cite{NdAlSiNeutron,NdAlSiQOchen}. A fit of $\kappa_{ph}$ using modified Callaway model considering a simplified two-level system for resonant scattering (Supplementary Note 2) can reproduce the $T$ dependence of experimental data (Fig.\,1f); the effective energy splitting is determined to be $\sim$40\,K (3.45 meV). According to the fit, an additional term in $\kappa_{xx}$ besides $\kappa_e$ and $\kappa_{ph}$ probably emerges below $\sim$20\,K (Fig.\,1f), consistent with a recent report \cite{NdAlSithermal}. We assign it to the contribution of magnetic excitations (i.e., magnons for $T < T_m$ and paramagnons/fluctuations \cite{RuCl3} for $T > T_m$).

More information about the influence of magnetic excitations on heat conduction is revealed by a magnetothermal conductivity study. In Fig.\,2a, we plotted $\kappa_{xx}(H)$ ($H \parallel c$) at different $T$. From 4\,K to 15\,K, the isotherms are nonmonotonic with minima occurring at an intermediate $H$; for $T\gtrsim$ 12 K, there is a single minimum located at $H_{min}$ in $\kappa_{xx}(H)$ (hollow triangles), whereas at lower $T$ two minima ($H_{min}^1$ and $H_{min}^2$, solid triangles) appear on each curve. We point out that the low-$H$ decrease of $\kappa_{xx}(H)$ can be mainly associated with $\kappa_e$ (short-dashed line in Fig.\,2a, see also Supplementary Fig.\,1), whereas the high-$H$ increase is most likely to be contributed by enhanced $\kappa_{ph}$ (the magnon thermal conductivity is supposed to decay with increasing $H$ \cite{NdAlSithermal}). The $T$ dependence of the characteristic fields determined from minima in $\kappa_{xx}(H)$ is displayed in Fig.\,2b, the implications of their possible origins are discussed in Supplementary Note 3. Note that $H_{min}$ roughly falls onto a line extrapolating to ($H, T$) = (0, 0) (magenta dashed line in Fig.\,2b), pointing towards a putative $H/T$ scaling behaviour. Such a scaling is clearly verified in Fig. 2c, where $\Delta \kappa_{ xx}/\kappa_{xx}{(\rm 0)}$=$[{\kappa_{xx}(H)}-{\kappa_{xx}({\rm 0})}]$/$\kappa_{xx}({\rm 0})$ is plotted as a function of $H/T$. For $T >$ 12\,K, the minimum of $\kappa_{xx}(H)$ occurs at a fixed $H/T$, a hallmark of the resonant phonon scattering \cite{Pr2Ir2O7} that reflects the Zeeman effect of the 4$f$ CEF levels. The $H/T$ scaling is deviated at $T <$ 12\,K for both $H_{min}^1$ and $H_{min}^2$. Below $\sim$ 8\,K, a hump-dip feature develops near $H_{min}^2$ with a local maximum showing up ahead of it (Fig.\,2a), which probably manifests a sudden change in the scattering rate of heat carriers in the vicinity of $H_p$ (Supplementary Note 3); above this feature, a remarkable increase of $\kappa_{xx}$ occurs and becomes stronger upon cooling. We suggest that while the high-$H$ upturn of $\kappa_{xx}(H)$ can be naturally understood as a result of the weakened resonant scattering between phonons and 4$f$ CFE levels, its strong enhancement at low $T$ suggests that another phonon scattering mechanism, i.e, the scattering off magnetic excitations (magnons and/or paramagnons), comes into play; a severe suppression of the latter process at high $H$ leads to an enhanced $\kappa_{ph}$ that can exceed 200$\%$ of the zero field $\kappa_{xx}$ at 4\,K (Supplementary Fig.\,1a).

The drastic depletion of scattering between phonons and (para)magnons strongly implies the development of a spectral gap for the latter. Similar to the case in quantum magnets $\alpha$-RuCl$_3$ \cite{RuCl3} and Na$_2$Co$_2$TeO$_6$ \cite{Na2Co2TeO6}, it is likely that low-energy spin excitations are gapped out in the ($H > H_p$) spin-polarized state of NdAlSi. This conclusion is supported by specific heat (Fig.\,2d) and ${^{27}}$Al nuclear magnetic resonance (NMR, inset of Fig.\,2e) measurements (see Methods for details). A fit of the magnetic heat capacity $C_m$ to a thermal activation model (Methods and Supplementary Fig.\,2) for 2\,K $< T <$ 10\,K yields a magnetic excitation gap $\Delta$ that opens roughly linearly with increasing $H$ \cite{NdAlSiQOchen} above 8\,T (Fig.\,2e). The gap opening field $H_g$ (extrapolated to $\Delta$ = 0) $\sim$ 3.5-4.0\,T appears to be lower than $H_m$ at low $T$ ($H_m \approx H_p \approx$ 5.5\,T for $T$ = 2\,K \cite{NdAlSiQO}). Fitting of the spin-lattice relaxation rate $1/T_1$ obtained from NMR experiment performed at $\mu_0H =$ 12\,T (Methods) also reveals a magnon gap consistent with the specific heat results (Fig.\,2e). The increase of $\Delta$ from $\sim$10-15\,K at 8\,T to $\sim$25-35\,K at 14\,T can thus elucidate the reduction of phonon scattering off the magnetic excitations in this field range. In all, our results demonstrate that the phonon heat conduction in NdAlSi is appreciably hindered by spin-phonon coupling, predominantly  scattering between phonon and magnetic excitations below $\sim$ 10\,K and resonant phonon scattering on local moments for higher $T$.

\bigskip
\noindent
\textbf{\normalsize Thermal Hall effect}

\noindent
Now we turn to the electronic heat transport and examine the impact of scattering of local spins on the Lorenz number. Because phonons (and magnons at low $T$) contribute to a large portion of $\kappa_{xx}$ and $\kappa_{zz}$ (Fig.\,1), determination of $L$ from the longitudinal transport is difficult. Here, we use the Hall Lorenz number $L_{xy} \equiv \frac{\kappa_{xy}}{\sigma_{xy} T}$ ($\kappa_{xy}$ and $\sigma_{xy}$ are the thermal and electrical Hall conductivities, respectively) to test the validity of WF law in NdAlSi. $L_{xy}$ also takes the value of $L_0$ if the WF law is obeyed; otherwise, considering that the transverse transport coefficients $\kappa_{xy}$ and $\sigma_{xy}$ contain square term of the scattering rate, it is proposed that $L_{xy}/L_0 \sim (L_{xx}/L_0)^2$ \cite{Cu-YBCO}. In the upper panels of Fig.\,3 we display the data of $\kappa_{xy}$ measured at various temperatures on a NdAlSi single crystal (sample \#1, see Methods and Supplementary Note 5 for experimental details), together with the hypothetic curves $\kappa_{xy}^{\rm WF} = L_0\sigma_{xy}T$ simulated based on the measured $\sigma_{xy}$. The $H$-dependent $L_{xy}$ determined from the ratio between $\kappa_{xy}$ and $\kappa_{xy}^{\rm {WF}}$ are plotted in the bottom panels of Figs.\,3. (More data taken on the same sample are shown in Supplementary Fig.\,3) Surprisingly, at $T$ = 4\,K, $\kappa_{xy}(H)$ appears to be larger than $\kappa_{xy}^{\rm WF}$ over a broad range of $H$ (Fig.\,3a): the value of $L_{xy}$ exceeds 4$\times$10$^{-8}$\,W $\Omega$ K$^{-2}$ (i.e., $L_{xy} >$ 1.6 $L_0$) across the up-down-down ferrimagnetic phase regime below $H_m$ and the field-induced paramagnetic regime between $H_m$ and $H_p$ \cite{NdAlSiQO} (vertical bars in Fig.\,3a); as the Nd 4$f$ moments become fully polarized above $H_p$, $L_{xy}(H)$ starts to decrease and the WF law is roughly recovered at $\mu_0H$ = 14\,T. For $T > T_m$ = 7.3\,K, the helical ferrimagnetic order is absent, yet the enhanced Hall Lorenz number $L_{xy}(H) > L_0$ over the entire field range (0 $\leq \mu_0H \leq $ 14\,T) persists up to $T \sim$ 30\,K. The maximum $L_{xy}$ reaches $\simeq$ 2\,$L_0$ at 8\,K and $\mu_0H$ = 5.5\,T (Fig. 3b). The $\kappa_{xy}(H)$ measured at 40\,K (Fig.\,3f) and even higher temperatures (Fig.\,3g,h and Supplementary Fig.\,3) exhibits only marginal deviations, i.e., within the experimental error (Supplementary Note 1), compared with $\kappa_{xy}^{\rm WF}$.

The thermal Hall study unveils an unusual enhancement of the Lorenz number in NdAlSi, suggesting a unique mechanism for the finite-$T$ WF law violation in this compound; such enhancement survives much higher temperature and fields than the ferrimagnetic order, thus could not be straightforwardly linked to a special magnetic configuration. Nevertheless, the enhancement appears to be suppressed once the spin polarization is established at high $H$ (Figs.\,3a-c). The overall behaviours have been reproduced in three more samples (\#2, \#3 and \#4), see Supplementary Fig.\,4. In all samples, an $L_{xy}$ considerably exceeding $L_0$ is observed at low $T$, with a maximum value of $\sim$ 1.6-2.4\,$L_0$. Due to the uncertainties in the measurement of sample dimensions and lead distances, there may be an error of 10-15\,$\%$ in the absolute values of $\kappa_{xy}$ and $\sigma_{xy}$, which precludes a detailed analysis of the potentially sample-dependent $L_{xy}$. Also, this error may contribute to the mismatches between the $\kappa_{xy}^{\rm WF}$ and measured $\kappa_{xy}$ at $T \gtrsim $100\,K (Supplementary Figs.\,3 and 4), making it difficult to comment on the exact validity of WF law for elevated temperatures (Supplementary Note 1). We note that the low-$T$ enhancement of $L_{xy}$ is absent in the isostructural compound SmAlSi, another Weyl semimetal with helical magnetic order \cite{SmAlSi}, in which the WF law is obeyed within the experimental error (see Supplementary Fig.\,5). In the nonmagnetic counterpart LaAlSi, $L_{xy}\simeq L_0$ is also verified (Supplementary Fig.\,6).

We further measured the thermal Hall conductivity for an in-plane $H$, i.e., $\kappa_{xz}$. Owing to the difficulty of applying $J_Q$ along $a$ and $c$ in the same sample, we determined $\kappa_{xz}$ by measuring transverse thermal gradient $\nabla T_z$ in the $ac$-plane (with $J_Q \parallel a$) and longitudinal $\kappa_{zz}$ in two different samples; $H$ was applied along $b$ in both cases (Supplementary Note 5). The results are summarized in Fig.\,4. The most notable feature on $\kappa_{xz}$ is a nonsaturating upward deviation from the WF value, $L_0\sigma_{xz}T$, with increasing $H$; the Lorenz number $L_{xz}$ (defined following the same way as $L_{xy}$) reaches $\sim$ 2.2\,$L_0$ at $T$ = 20\,K and $\mu_0H$ = 14\,T (Fig.\,4b). Such enhanced $L_{xz}$ is still evident at least up to 50\,K (Fig.\,4c), whereas for $T \gtrsim$ 100\,K the potential deviation from $L_0$ is elusive due to experimental uncertainties (Figs.\,4d-f). The nonmonotonic $H$-dependence manifested on the low-$T$ $L_{xy}$ (Figs.\,3a-d) is absent on $L_{xz}$. We suggest that this phenomenon is probably linked to the high saturation field for magnetization with $H$ applied in-plane (Supplementary Note\,7 and Supplementary Fig.\,9f; for $H \parallel b$, full spin polarization is unlikely to be achieved below 14\,T), consistent with the easy-axis anisotropy of NdAlSi \cite{NdAlSiNeutron,NdAlSiChen}. Hence, the comparison provides further evidence that the high-$H$ drop of $L_{xy}$ is associated with the spin-polarized state.

\bigskip
\noindent
{\textbf{\large Discussion}}

\noindent
It has been verified that in various metals and semimetals \cite{BehniaWP2,BehniaAntimony,Cu-YBCO,CeRhIn5}, abundance of inelastic $e-e$ and $e-ph$ scattering explicitly leads to a downward deviation from the WF law (i.e., $L < L_0$) at intermediate temperatures. This is because the entropy flow is impeded by all types of inelastic scattering while the charge current is degraded only due to the momentum-relaxing (large-angle) scattering processes \cite{PrincipiPRL,BehniaAntimony}. Our observation of $L_{xy} > L_0$ in NdAlSi is, however, in contrary to such a universal picture elucidating a reduction of $L$, implying the presence of unique underlying physics that may have been overlooked previously. At first glance, the abnormally enhanced $L_{xy}$ may reflect the inclusion of a nonelectronic contribution in $\kappa_{xy}$. Despite that charge-neutral excitations such as phonons \cite{Pr2Ir2O7,Tb3Ga5O12,STOthermalHall,Cu3TeO6,BPthermalHall} and magnons \cite{Onose,Cubdc} can give rise to a thermal Hall signal as well, we have sufficient evidence to rule out them as the origin of the enhanced $L_{xy}$ in NdAlSi. The $T$ dependence of $\kappa_{xy}$ shown in Fig.\,5a indicates that the excess thermal Hall (exceeding the WF value given by the solid line) is most prominent between $\sim$ 5-25\,K. The lack of corresponding feature in $\kappa_{xx}$ (Fig.\,1d) in this temperature range strongly disfavors the phonon Hall scenario \cite{STOthermalHall}; the fact that the excess $\kappa_{xy}$ persists up to at least 4$\times T_m$ ($\approx$ 30\,K, Fig.\,5b) also makes the magnon origin highly unlikely. Moreover, the amplitude of excess thermal Hall in NdAlSi invalidates the interpretations of phonon Hall and magnon Hall: it easily reaches $(4-5)\times10^2$ mW m$^{-1}$K$^{-1}$ (Fig.\,3, Supplementary Figs.\,3 and 4), composing about one half of the total thermal Hall angle $\mid \kappa_{xy}/\kappa_{xx} \mid \simeq$ 35$\%$ at 5.5\,T and $T \simeq$ 10-12\,K (Fig.\,5a). In contrast, a universal upper limit of $\mid \kappa_{xy}/(\mu_0 H \kappa_{xx}) \mid <$ 2$\times$10$^{-3}$\,T$^{-1}$ has been revealed for the phonon Hall angle \cite{Cu3TeO6,BPthermalHall,Sr2IrO4thermalHall} (see Supplementary Note 10 and Supplementary Table 1 for details); the magnon Hall is characterized by an even smaller Hall angle of 10$^{-3}$ \cite{Onose}.

Having established the electronic nature of $L_{xy}$ enhancement, we can further narrow down the speculations on its cause. The anomalous transverse terms stemming from the Berry curvature of electron wavefunctions \cite{Mn3Ge} are irrelevant, because NdAlSi does not exhibit noticeable anomalous Hall effect (Supplementary Note 6 and Supplementary Fig.\,10d) \cite{NdAlGe}. In doped SrTiO$_3$, the electronic thermal Hall can be significantly boosted by a magneto-thermoelectric effect, which introduces an additional term $\Delta\kappa_{xy} \sim -\alpha_{xy}S_{\rm drag} T$ ($\alpha_{xy}$ and $S_{\rm drag}$ are the transverse thermoelectric conductivity and the phonon-drag term in thermopower, respectively) \cite{STOPhononDrag}. For NdAlSi, however, such a scenario is dismissed considering the vanishingly small $S_{\rm drag}$ (Supplementary Fig.\,7a). A drastically increased $L$ up to $\sim$ 20 $L_0$ has been reported in graphene near its charge-neutrality point \cite{graphene}; the involved exotic heat carriers, i.e., the neutral Dirac fermion fluid, shall not leave any fingerprints in the Hall channel.

A model proposed by Goff over half a century ago, which was aimed at explaining the anomalous $L > L_0$ in chromium \cite{CrGoff1,CrGoff2}, may help us understand the enhanced ratio of heat to charge transport in NdAlSi. In Goff's model, a minimum centered at $\epsilon = E-\mu$ = 0 ($\mu$ is the chemical potential) characterizes the energy dependence of specific conductivity, $\sigma(\epsilon) \propto N(\epsilon)v^2(\epsilon)\tau(\epsilon)$, where $N(\epsilon)$ is the density of states (DOS), $v$ is the carrier group velocity and $\tau$ is the relaxation time. Such a dip-like profile results in a weaker suppression of $\kappa$ compared with $\sigma$ at finite $T$, leading to an increase of $L$ (see Supplementary Note 8 for details). For chromium, the dip feature is associated with the spin-density-wave gap that erases the electronic DOS on parts of its Fermi surfaces \cite{CrGoff2}. Later, the same model was invoked to interpret the enhanced $L$ in high-temperature superconductors wherein the pseudogap (PG) physics is involved \cite{EuBaCuO,YBCOMatusiak,BaFe1-xCoxAs} (the PG corresponds to a depletion of DOS centered at $\mu$). For NdAlSi, there is no evidence for such PG-like DOS depletion. Instead, we attribute the large $L_{xy}$ to a depression of $\tau(\epsilon)$ that is symmetrical about $\epsilon$ = 0, as shown in Fig.\,5c, which equally renders a dip in $\sigma(\epsilon)$. Taking such an attenuated relaxation time into account, the integrands of $\kappa_{xy}$ ($Q$) and $\sigma_{xy}$ ($P$) (Supplementary Note 8) show distinct energy profiles (Fig.\,5c): at finite $T$, the former has more states contributing to the energy integral, giving rise to excess thermal Hall compared with electrical Hall, i.e., an enhanced $L_{xy}$.

In NdAlSi, the peculiar energy dependence of $\tau$ is most likely due to the elastic Kondo-type scattering between itinerant (Nd 5$d$) and localized (Nd 4$f$) electrons. It is known that Kondo scattering process can lead to such profile of $\tau(\epsilon)$ and, subsequently, an increased $L$ \cite{LaCeAl2,CeAl3,CeCu6}. The most prominent interaction involving 5$d$ and 4$f$ electrons in this material is the RKKY interaction, as the itinerant Weyl fermions mediate the indirect exchange coupling between 4$f$ moments \cite{NdAlSiNeutron}. Nonetheless, we point out that Kondo-type scattering is likely to coexist with the RKKY coupling: in Doniach's phase diagram \cite{Doniach}, a system containing itinerant electrons and local moments in the weak-coupling and low-electron-DOS limits (which is the case of NdAlSi) is characterized by the coexistence of Kondo interaction and RKKY interaction, whilst the latter prevails (Fig.\,5d). By considering higher-order scattering in an incoherent Weyl-Kondo scattering model (Supplementary Note 9), we can reproduce the energy-symmetric minimum of $\tau(\epsilon)$ at $\epsilon$ = 0 (Fig.\,5c); a numerical calculation based on the resulted integrand functions $Q$ and $P$ (Supplementary Note 9) yields a $T$-dependent $L_{xy}$ that satisfyingly tracks the trend of the experimental data (red line in Fig.\,5b). In this scenario, the high-$H$ reduction of $L_{xy}$ (Figs.\,3a-c) can be naturally linked to weakened spin-flip scattering in the fully polarized state, wherein 4$f$ moments are strongly pinned by the external field. The absence of notable excess thermal Hall in SmAlSi (Fig.\,5b, Supplementary Fig.\,5) probably reflects a much weaker Kondo scattering therein, whereas the validity of WF law in nonmagnetic LaAlSi (Supplementary Fig.\,9) is fully consistent with our expectation.

The assignment of the underlying mechanism for the enhanced $L$ to Kondo-type spin-flip scattering finds support from a recent experimental work, which reveals Kondo hybridization between local moments and Weyl fermions in another magnetic topological semimetal, CeAlGe \cite{CeAlGeMingdaLi}. Similar Kondo physics may come into play in a broad range of magnetic topological materials (note that the qualitative behaviour of Kondo-scattering-enhanced $L_{xy}(T)$ shown in Fig.\,5b does not rely on the position of band crossing, nor does it change significantly when parabolic bands exist at $\mu$, see Supplementary Note 9 and Supplementary Fig.\,11); further explorations are required to elucidate its impact on various physical properties. We also note that an unconventional Nernst effect has recently been resolved in NdAlSi and is proposed to arise from a remarkable energy dependence of carrier relaxation time \cite{NdAlSiNernst}; such energy dependence is predominantly linked to the abundance of scattering from magnetic fluctuations at the ``hot spots" on specific Fermi surfaces, triggered by a Fermi surface nesting instability. The same mechanism can contribute to thermal transport as well and possibly adds to the enhancement of $L_{xy}$. Nevertheless, the impact of magnetic fluctuation scattering is rapidly suppressed in the magnetically ordered state \cite{NdAlSiNernst} and thus the large excess thermal Hall in the u-d-d state (Fig.\,3a) cannot be attributed to this origin. At the present stage, the incoherent Kondo-type scattering model serves as the most applicable interpretation of experimental observations, though a more complete understanding may rely on future investigations invoking microscopic probes.

Our results unequivocally verify the intricate and enriched coupling between low-energy excitations and local magnetism in the magnetic Weyl semimetal NdAlSi. The phonon conduction is shown to be strongly affected by resonant scattering on CEF-split 4$f$ multiplets and, at lower $T$, scattering off magnetic excitations. Moreover, apart from the Weyl-mediated RKKY interaction, we suggest that there is also a subtle Kondo-type interaction between the Weyl fermions and the local moments; such Kondo-like scattering endows the relaxation time with a unique energy dependence, which leads to an unusual deviation from the WF law with a remarkably enhanced Lorenz number. The relevance of Kondo physics in NdAlSi highlights the emergent correlation effects stemming from the interplay between local spins and relativistic quasiparticles in topological materials. On a final note, our work demonstrates that thermal Hall measurement, combining with the WF analysis, is a powerful probe for unveiling special energy dependence of transport coefficients within a narrow energy window (a few meV) around the Fermi energy. Further development of this technique may help clarify the novel physics in numerous correlated quantum materials.

\bigskip

\noindent
{\bf \large {Methods}}\\
\noindent
{\bf Single crystal growth and characterizations}
\noindent
We synthesized high-quality single crystals of NdAlSi, LaAlSi and SmAlSi using the Al self-flux method \cite{NdAlSiQO}. The crystal orientation and stoichiometry were determined by the pattern of single-crystal X-ray diffraction (XRD) and  energy dispersive X-ray spectroscopy (EDX), respectively (Supplementary Fig.\,8). Electrical transport measurements were performed using conventional Hall-bar configuration in a Quantum Design DynaCool-14\,T Physical Property Measurement System (Supplementary Fig.\,10, see Supplementary Note 6 for details). Magnetic properties were measured by a Quantum Design vibrating sample magnetometer (VSM) up to 7\,T (Supplementary Fig.\,9, see Supplementary Note 7 for detailed discussions). In this work we studied six pieces of NdAlSi single crystals (samples \#1, \#2, \#3, \#4, \#5 and \#6); they were grown in the same batch and all cut and polished into a rectangular bar shape before measurement. The dimensions (length$\times$width$\times$thickness) of the samples were 1.8$\times$1.0$\times$0.25\,mm$^3$ (\#1), 2.0$\times$0.9$\times$0.19\,mm$^3$ (\#2), 2.6$\times$1.1$\times$0.21\,mm$^3$ (\#3), 2.6$\times$0.9$\times$0.19 \,${\rm {mm}}^3$ (\#4), 0.69$\times$0.58$\times$0.50\,mm$^3$ (\#5) and 2.6$\times$1.1 $\times$0.57\,mm$^3$ (\#6). For samples \#1-\#4, measurements of $\kappa_{xx}$ and $\kappa_{xy}$ were performed with heat current $J_{\rm Q} \parallel a$ and $H \parallel c$, whereas $\kappa_{xz}$ was determined from the data measured in samples \#5 and \#6 (Supplementary Note 5). Here, the coordinates $x$, $y$, and $z$ correspond to the crystallographic $a$([100]), $b$([010]) and $c$([001]) axes, respectively.

\bigskip

\noindent
{\bf Thermal transport and thermoelectric measurements}
\noindent
In our thermal transport experiment, the sample was mounted on a home-built thermoelectric puck, which was loaded into a Quantum Design physical property measurement system (PPMS-9\,T or 14\,T). Measurements were taken under high vacuum condition (pressure lower than 1$\times 10^{-5}$ Torr). We used a one-heater-two-(differential-)thermocouple setup as shown in Fig.\,1c, which allows the simultaneous measurement of the longitudinal and transverse temperature differences ($\Delta T_x$ and $\Delta T_y$). A 1\,k$\Omega$ chip resistor was attached to one end of the sample by VGE varnish serving as the heater; a low frequency AC current was applied to the heater by a Keithley 6221 current source. The quasi-AC method was adopted to reduce the DC background noise which appears to be inevitable in PPMS. The other end of the sample was fixed to a copper block with a piece of cigarette paper soaked by diluted VGE-varnish serving as the insulating layer. Two pairs of type-E thermocouple were attached to the sample by stycast 2850\,FT (with carefully ensured electrical insulation between the thermocouple and sample) in a differential configuration depicted in Fig.\,1c. The voltage signals on thermocouple were collected by Keithley 2182A nanovoltmeters. We monitored $J_Q$ to guarantee that $\Delta T_x$ is smaller than 10$\%$ of the base temperature at $T < 5\,K$ and $\Delta T_x < 0.5\,K$ for $T > 5\,K$. Using the same setup, the longitudinal (Seebeck) and transverse (Nernst) thermoelectric coefficients can be also measured by attaching two pairs of gold wires to the sample with silver paste. For the field-sweep measurements, the assembly was thermally stabilized for at least 15 minutes before field ramping at each temperature point. The sweeping rate of magnetic field was set as 2\,mT/s. Additional thermal transport measurements utilizing steady-state methods and thermometers (instead of thermocouples) were also performed for a double check of the main results; for details of these examinations, see Supplementary Note 1.

\bigskip

\noindent
{\bf Specific heat measurements and data analysis}
\noindent
Specific heat was measured by a long relaxation method in a Dynacool-14\,T PPMS (Quantum Design). As shown in Fig.\,2d and Supplementary Fig.\,2a, at zero field, $C(T)$ exhibits a sharp $\lambda$-type peak at $T_m$ = 7.3\,K (onset of ferrimagnetic order) and a weaker feature at $T_{com} \simeq$ 3.3\,K (an incommensurate-to-commensurate magnetic transition), as well as a broad anomaly at around 16\,K which is attributed to a CEF Schottky term $C_{Sch}$ \cite{NdAlSiNeutron,NdAlSiQOchen}. With increasing $H$, the two magnetic transitions are gradually smeared out and the Schottky hump moves to higher $T$. $C_{Sch}$ can be well reproduced by assuming a ground state doublet separated from four degenerate excited doublets for the Kramers ion Nd$^{3+}$ (the sketch depicted in the inset of  Supplementary Fig.\,2c), i.e., $C_{Sch} \propto {\Delta_s^2}/{[\cosh({{\Delta_s}/{2k_{\rm B}T}})}]^2$, where $\Delta_s$ is a CEF splitting energy of $\sim$ 4.1 meV \cite{NdAlSiQOchen}. Such picture is consistent with the suppressed phonon thermal conductivity in NdAlSi (Fig.\,1e) stemming from the resonant scattering on the Nd$^{3+}$ CEF levels. 

The magnetic heat capacity $C_m$ of NdAlSi can be obtained by a subtraction of (predominantly phonon) background estimated from the specific heat of nonmagnetic counterpart LaAlSi (Supplementary Fig.\,2b). A more detailed analysis of the phonon contribution to specific heat is presented in Supplementary Note 4. Apart from the Schottky term $C_{Sch}$ discussed above (red line in Supplementary Fig.\,2b, inset), $C_m$ contains the contribution from a system with independent magnetic excitations, which dominates at low $T$ and can be described as \cite{URu2Si2Capacity}:
\begin{equation}
C_m = \int_{0}^{\infty}Eg(E)\frac {df(E)}{dT}dE,
\end{equation}
where $g(E)$ is the density of states (DOS) and $f(E)=1/[{\rm exp}({E/k_BT}-1)]$ is the Bose factor for magnetic excitations. For an isotropic excitation spectrum taking the form $E=\Delta_{\varepsilon} +\alpha k^{\beta}$, $g(E)=(V_m/2\pi^2\beta \alpha ^{3/\beta})(E-\Delta_{\varepsilon})^{3/\beta-1}$. Thus, the acoustic branches as the leading term contributing to specific heat gives $C_m = AT^{\rm n}{\rm exp}(-\Delta/T)$ (here $n=3/\beta-2$ and $\Delta = \Delta_{\varepsilon}/k_B$), which is a thermally activated form controlled by a magnetic excitation gap $\Delta$. For NdAlSi, the specific dispersion relation (i.e., $\beta$) of magnon band remains unknown, thus we applied this model to our data below 10\,K with two fixed values of the exponent $n$:  $n$ = 0 ($\beta$ = 3/2) and $n$ = -1 ($\beta$ = 3). The two choices yield slightly different $\Delta(H)$ (Fig.\,2c). Nevertheless, the main conclusion that a magnon gap is opened linearly by a magnetic field $\mu_0H (\parallel c) \gtrsim$ 3.5\,T stays valid.
\bigskip

\noindent
{\bf Nuclear magnetic resonance measurements}
\noindent
Nuclear magnetic resonance (NMR) measurements were performed using a commercial NMR spectrometer from Thamway Co.Ltd in an NMR-quality 12\,T magnet (Oxford Instruments). Data were taken under an external magnetic field of 12\,T along the $c$ direction. The nuclear spin-lattice relaxation time $T_1$ of $^{27}Al$ nuclei was measured by the saturation-recovery method and the recovery of nuclear magnetization $M(t)$ was fitted by a function $1-M(t)/M(\infty)=I_0  \{0.028 \exp [-(t/T_1)^\beta]+0.178 \exp[-(6t/T_1)^\beta]+0.794 \exp[-(15t/T_1)^{\beta}]\}$. For the long wave approximation in antiferromagnetic materials, $1/T_1$ is given by $1/T_1 \propto T^3 \int_ {T_{\rm {AE}}/T}^ {\infty}xdx/(e^{x}-1)$, where $T_{\rm AE}$ represents the magnon excitation gap. For $T\ll T_{\rm {AE}}$, $1/T_1 \propto \,T^2$ $\exp(-T_{\rm AE}/T)$ \cite{NMR1,NMR2}. As shown in the inset of Fig.\,2e, $1/T_1$ is nearly $T$ independent above $\sim$ 10\,K as expected for a paramagnetic phase. Below 10\,K, the rapid decrease of $1/T_1$ indicates a pronounced depletion of spectral weight and the opening of an energy gap in the magnon excitation spectrum; the gap size given by $1/T_1 \propto \, T^2 \exp(-\Delta/T)$ is $\Delta \sim$ 29\,K.

\bigskip

\noindent
\textbf{\large Data availability}

\noindent
The experimental data presented in the main text figures have been deposited in the figshare database under accession code \href{https://doi.org/10.6084/m9.figshare.27336996}{https://doi.org/10.6084/m9.figshare.27336996}. More data are available from the corresponding authors upon request.

\bigskip

\noindent
\textbf{\large Code availability}

\noindent
The codes used for the fitting and simulation process in this study are available from the corresponding author upon request.

\bigskip

\noindent
\textbf{\large Acknowledgements}

\noindent
We thank Ziqiang Wang, Qian Niu, Zhenyu Wang and Shiyan Li for insightful discussions. We are grateful to Shuangkui Guang and Xuefeng Sun for valuable suggestions on temperature calibration. This work was financially supported by the National Key R$\&$D Program of the MOST of China (Grant No. 2022YFA1602602) (Z.X.), the Fundamental Research Funds for the Central Universities (Grant No. WK3510000014) (Z.X.), the Innovation Program for Quantum Science and Technology (2021ZD0302802) (T.W., Z.X. and X.-H.C.), the National Natural Science Foundation of China (Grants Nos. 12274390 (Z.X.) and 12174394 (J.Z.)), the Anhui Initiative in Quantum Information Technologies (AHY160000) (T.W. and X.-H.C.), the Anhui Provincial Major S$\&$T Project (s202305a12020005) (J.Z.), the High Magnetic Field Laboratory of Anhui Province under Contract (No. AHHM-FX-2020-02) (J.Z.), the HFIPS Director’s Fund (Grants Nos. YZJJQY202304 and BJPY2023B05) (J.Z.) and Chinese Academy of Sciences under contract No. JZHKYPT-2021-08 (T.W., J.Z., Z.X. and X.-H.C.). Z.X. acknowledges the USTC startup fund.

\bigskip

\noindent
\textbf{\large Author contributions}

\noindent
Z.X. and N.Z. proposed and designed the research. N.Z. grew the single crystals and performed the thermal transport, specific heat, electrical transport and thermoelectric measurements with help from H.-P.L, H.-Y.L., K.T. and T.Q.; N.Z. and Z.X. analyzed the data. N.Z. and K.T. performed the calibration for the thermal-transport assembly. L.N. and T.W performed the NMR measurements and relevant data analysis. D.T., D.L. and J.Z. performed the theoretical analysis and calculations. X.-H.C supervised the research project. N.Z. and Z.X. wrote the paper with inputs from all authors.

\bigskip

\noindent
\textbf{\large Competing interests}

\noindent
The authors declare no competing interests.

\bigskip
\bigskip

\newpage

\begin{figure}[h]
	\begin{center}
		\includegraphics[width=0.95\linewidth]{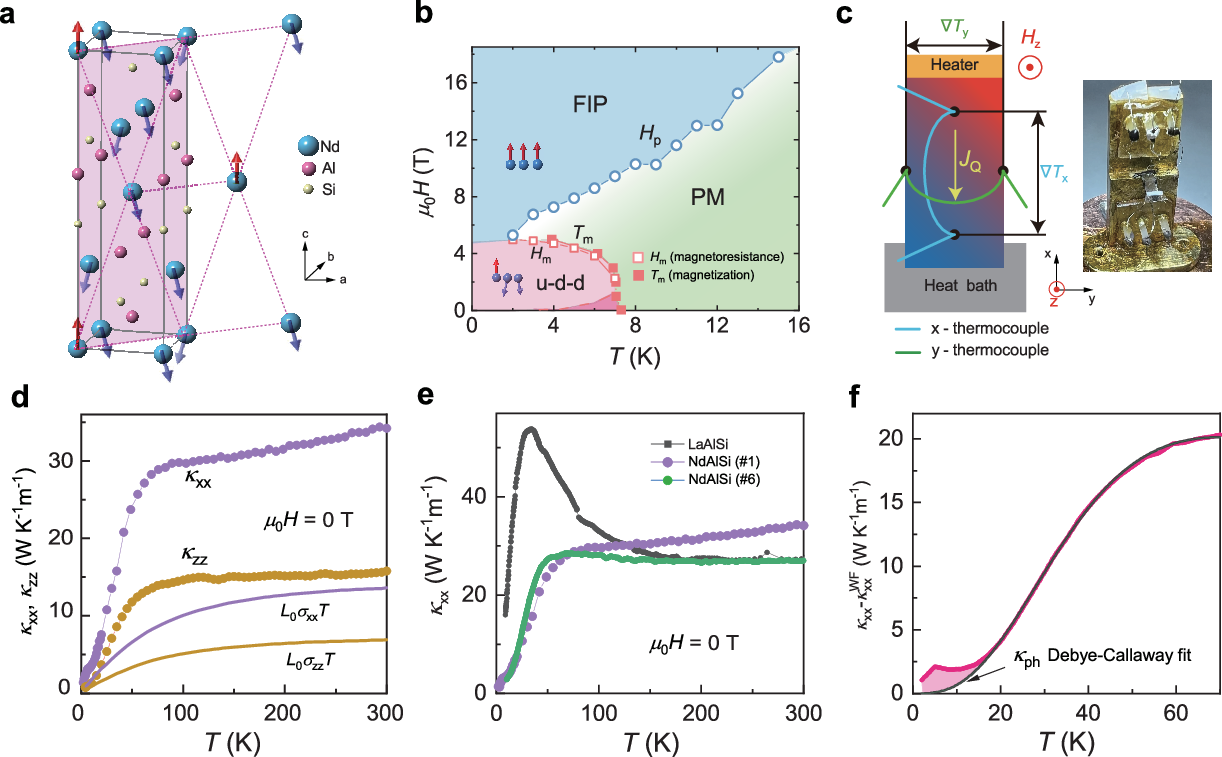}
		\caption{{\bf Phase diagram and the longitudinal thermal conductivity $\kappa_{xx}$ of NdAlSi.} {\bf a,} Crystal structure of NdAlSi. {\bf b,} The magnetic phase diagram of NdAlSi, showing the paramagnetic (PM), up-down-down (u-d-d) helical ferrimagnetic and field-induced polarized (FIP) states \cite{NdAlSiQO,NdAlSiChen,NdAlSiQOchen}. $H_m$ and $H_p$ are the metamagnetic transition field and the crossover field for spin polarization, respectively. {\bf c,} A schematic of the experimental setup for thermal transport measurements with two differential thermocouples (left); a photograph showing the thermoelectric puck with sample mount (right). {\bf d,} Temperature ($T$)-dependent longitudinal thermal conductivities $\kappa_{xx}$ (purple) and $\kappa_{zz}$ (yellow) of NdAlSi, measured with thermal gradient $\nabla T$ parallel to the crystallographic $a$ and $c$ axes, respectively. Solid lines are estimated electronic thermal conductivities assuming the validity of the WF law in the whole range of $T$. {\bf e,} Zero-field $\kappa_{xx}(T)$ measured in two NdAlSi samples (\#1 and \#6, purple and green circles, respectively) and a LaAlSi sample (black squares) from 2 to 300 K with $J_Q \parallel a$. {\bf f,} The estimated non-electronic contribution of $\kappa_{xx}$ in NdAlSi attained by subtracting the Wiedemann-Franz term $\kappa_{xx}^{\rm WF} = L_0\sigma_{xx}T$ from the total $\kappa_{xx}$. The black curve is a fit of the phonon thermal conductivity $\kappa_{ph}$ to a modified Debye--Callaway model (Supplementary Note 2).}
		\label{fig:figure1}
	\end{center}
\end{figure}

\begin{figure}[hbtp]
\begin{center}
	\includegraphics[width=0.95 \linewidth]{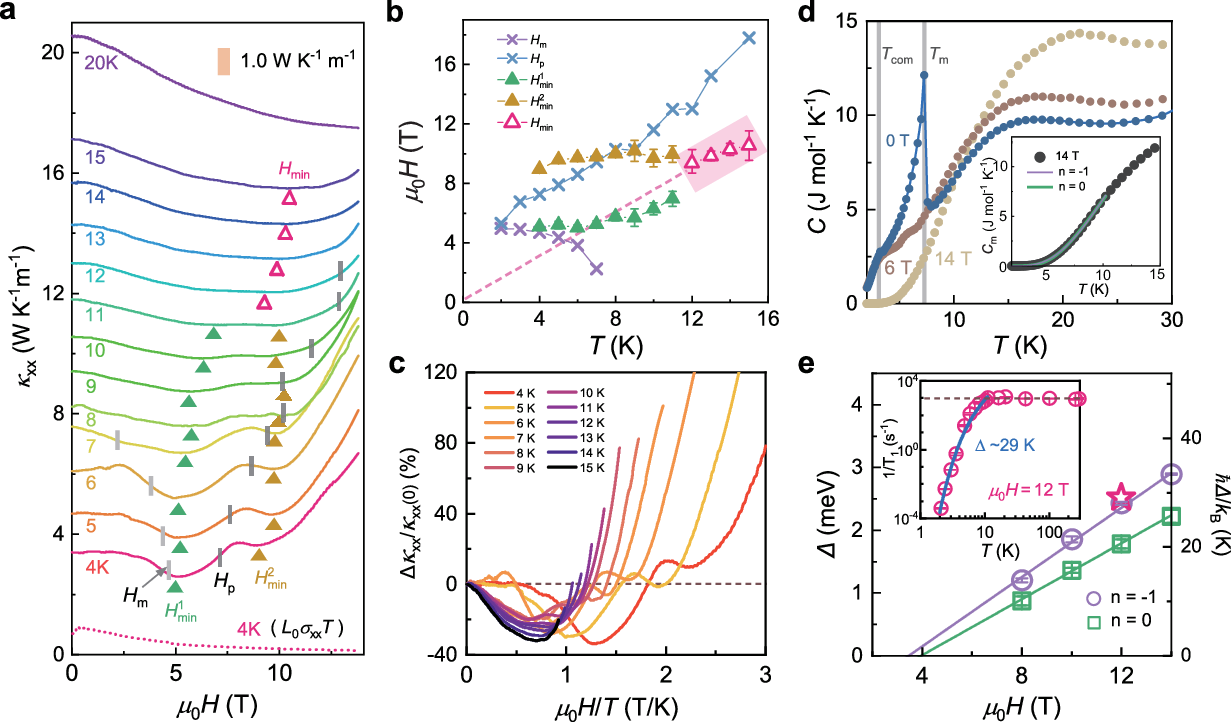}
	\caption{{\bf Longitudinal thermal conductivity and magnetic excitation gap of NdAlSi.} {\bf a,} $\kappa_{xx}$ measured at various temperatures plotted as a function of $H$ (applied along $c$). Curves are vertically shifted by 1 W m$^{-1}$K$^{-1}$ for clarity. Short-dashed line denotes the electronic contribution estimated based on the WF law at $T$ = 4\,K. Characteristic fields $H_{min}$ ($T$ $\geq$12 K) and $H_{min}^1$, $H_{min}^2$ ($T$ $\textless$ 12 K) are marked by hollow and solid triangles, respectively. $H_m$ and $H_p$ (see Fig.\,1b) are indicated by black and grey bars, respectively. {\bf b,} $T$ dependence of the characteristic fields defined in {\bf a}. Magenta thick line indicates a quasi-linear $H_{min} - T$ relationship extrapolating to $T$ = 0, $H$ = 0 (dashed line). {\bf c,} Magnetothermal conductivity normalized to the zero-field value, $\Delta \kappa_{xx}(H)/\kappa_{xx}(0)=[\kappa_{xx}(H) - \kappa_{xx}(0)]/\kappa_{xx}(0)]$, plotted against $H/T$. Error bars in {\bf b} are defined as the half widths of the field range wherein in the $H$ derivative of $\Delta \kappa_{xx}(H)/\kappa_{xx}(0)$ changes its value by 1. {\bf d,} Specific heat of NdAlSi measured under $\mu_0H$ = 0\,T (blue), 6\,T (brown) and 14\,T (yellow). The two vertical lines denote the magnetic transitions at $T_{\rm m}$ = 7.3 K and $T_{\rm {com}}$ = 3.3 K; the latter marks an incommensurate-to-commensurate transition inside the ferrimagnetic state \cite{NdAlSiNeutron}. Inset: Best fits of magnetic heat capacity $C_m$ at 14\,T to a thermal activation model (see Methods). Fits were applied to a temperature range 2 K $< T<$ 10 K. {\bf e,} The magnon excitation gap $\Delta$ (left axis) and corresponding temperature scale $\hbar\Delta/k_{\rm B}$ (right axis) determined from fits of $C_m(T)$ with the exponent $n$ = 0 (green squares), $n$ = -1 (purple circles), and analysis of the spin-lattice relaxation time $T_1$ (red star). The evolution of $\Delta$ under magnetic field is approximately linear with an gap opening field 3.5-4\,T. Inset: $T_1^{-1}$ plotted against $T$ for $\mu_0H$ = 12\,T. The solid line is a fit to an exponential function yielding the magnon gap (Methods).}
		\label{fig:figure1}
	\end{center}
\end{figure}

\begin{figure}[hbtp]
	\begin{center}
		\includegraphics[width=1.0\linewidth]{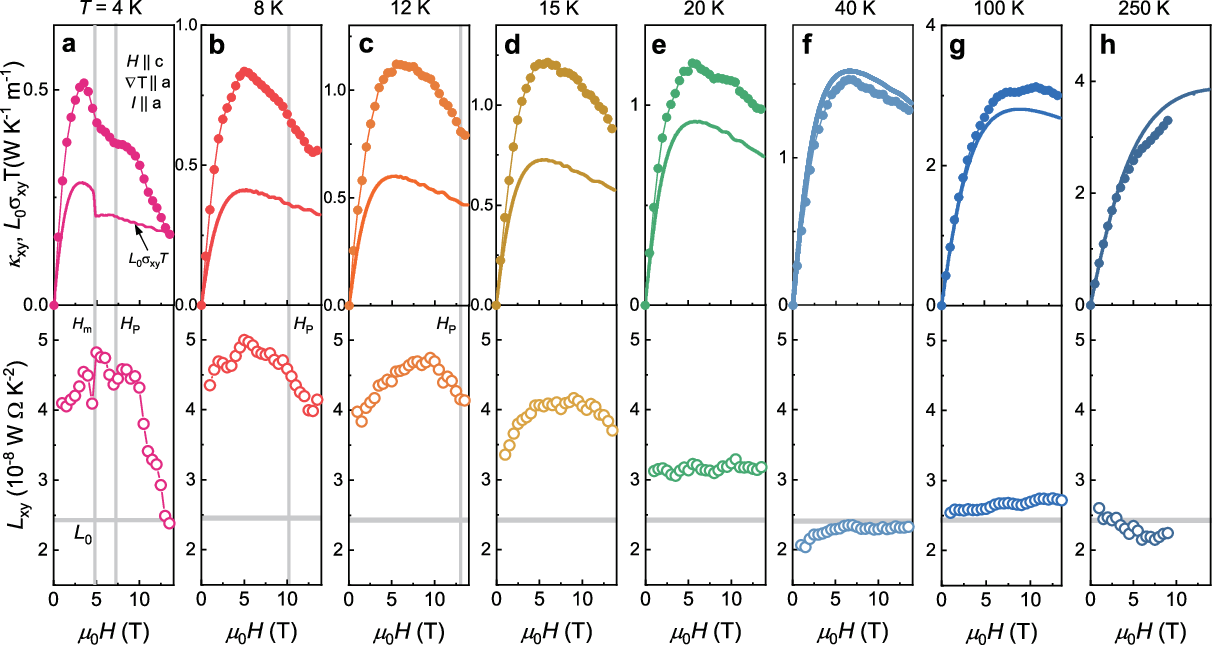}
		\caption{{\bf Magnetic field ($H$) dependence of the thermal Hall $\kappa_{xy}$ and the Hall Lorenz number $L_{xy}$ of NdAlSi.} Data were measured in sample \#1 with the temperature gradient $\nabla T$ and the $H$-field applied along the $a$ and $c$ axes, respectively. Top panels: Experimental $\kappa_{xy}(H)$ (solid circles) compared with the electronic contribution $\kappa_{xy}^{\rm WF} = L_0\sigma_{xy}T$ (solid lines) estimated based on a valid WF law. Vertical lines mark out the metamagnetic transition field $H_m$ and the crossover field $H_p$ for spin polarization \cite{NdAlSiQO}. Bottom panels:\,The $L_{xy}$ corresponding to the data shown in {\bf a-h}, plotted against $H$. The horizontal lines are the Sommerfeld value $L_0$. {\bf a,} $T$ = 4\,K, {\bf b,} $T$ = 8\,K, {\bf c,} $T$ = 12\,K, {\bf d,} $T$ = 15\,K, {\bf e,} $T$ = 20\,K, {\bf f,} $T$ = 40\,K, {\bf g,} $T$ = 100\,K, {\bf h,} $T$ = 250\,K.}
		\label{fig:figure2}
	\end{center}
\end{figure}

\begin{figure}[hbtp]
	\begin{center}
		\includegraphics[width=0.95\linewidth]{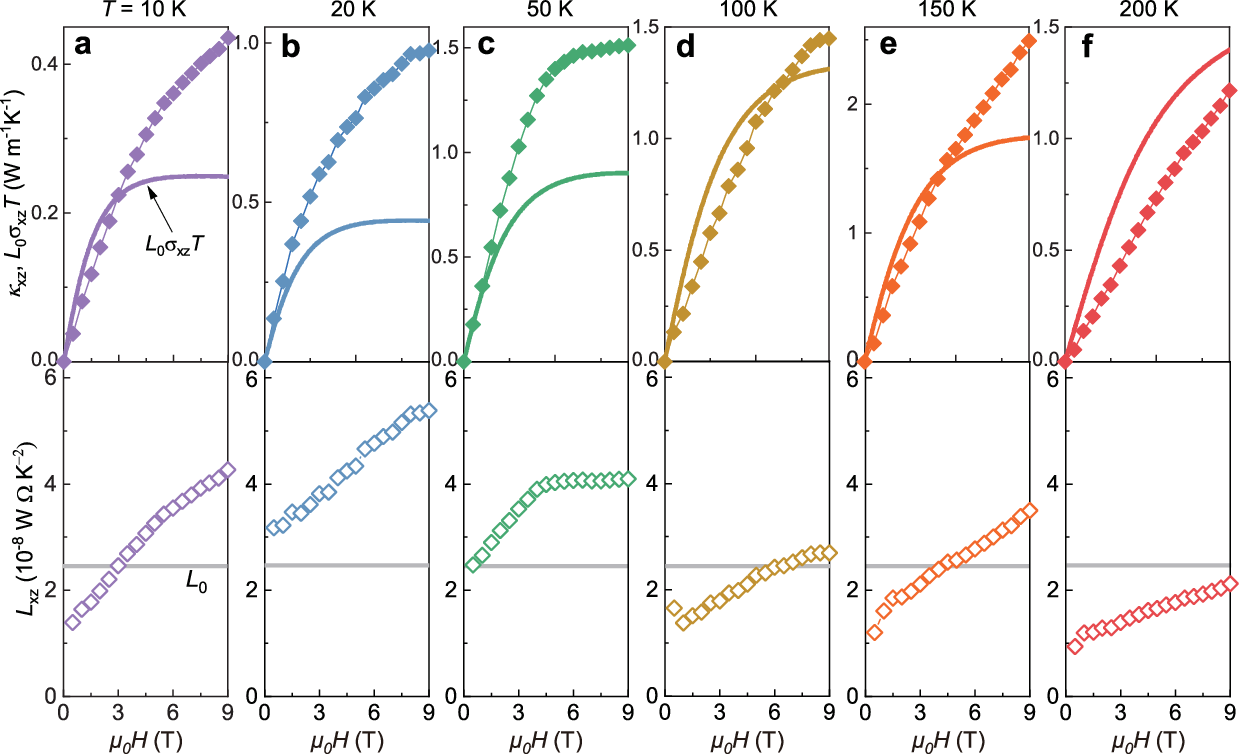}
		\caption{{\bf The out-of-plane thermal Hall $\kappa_{xz}(H)$ and $L_{xz}(H)$ of NdAlSi.} Data were measured with $H \parallel [010] (b)$ (see Supplementary Note 5 for details of the derivation of $\kappa_{xz}$.) Top panels: Experimental $\kappa_{xz}(H)$ (solid diamonds) compared with the electronic contribution $\kappa_{xz}^{\rm WF} = L_0\sigma_{xz}T$ (solid lines) estimated based on a valid WF law. Bottom panels: $L_{xz}(H)$ determined from $\kappa_{xz}(H)$ . The horizontal lines denote the Sommerfeld value $L_0$. {\bf a,} $T$ = 10\,K, {\bf b,} $T$ = 20\,K, {\bf c,} $T$ = 50\,K, {\bf d,} $T$ = 100\,K, {\bf e,} $T$ = 150\,K, {\bf f,} $T$ = 200\,K.}
		\label{fig:figure3}
	\end{center}
\end{figure}

\begin{figure}[hbtp]
	\begin{center}
		\includegraphics[width=0.85\linewidth]{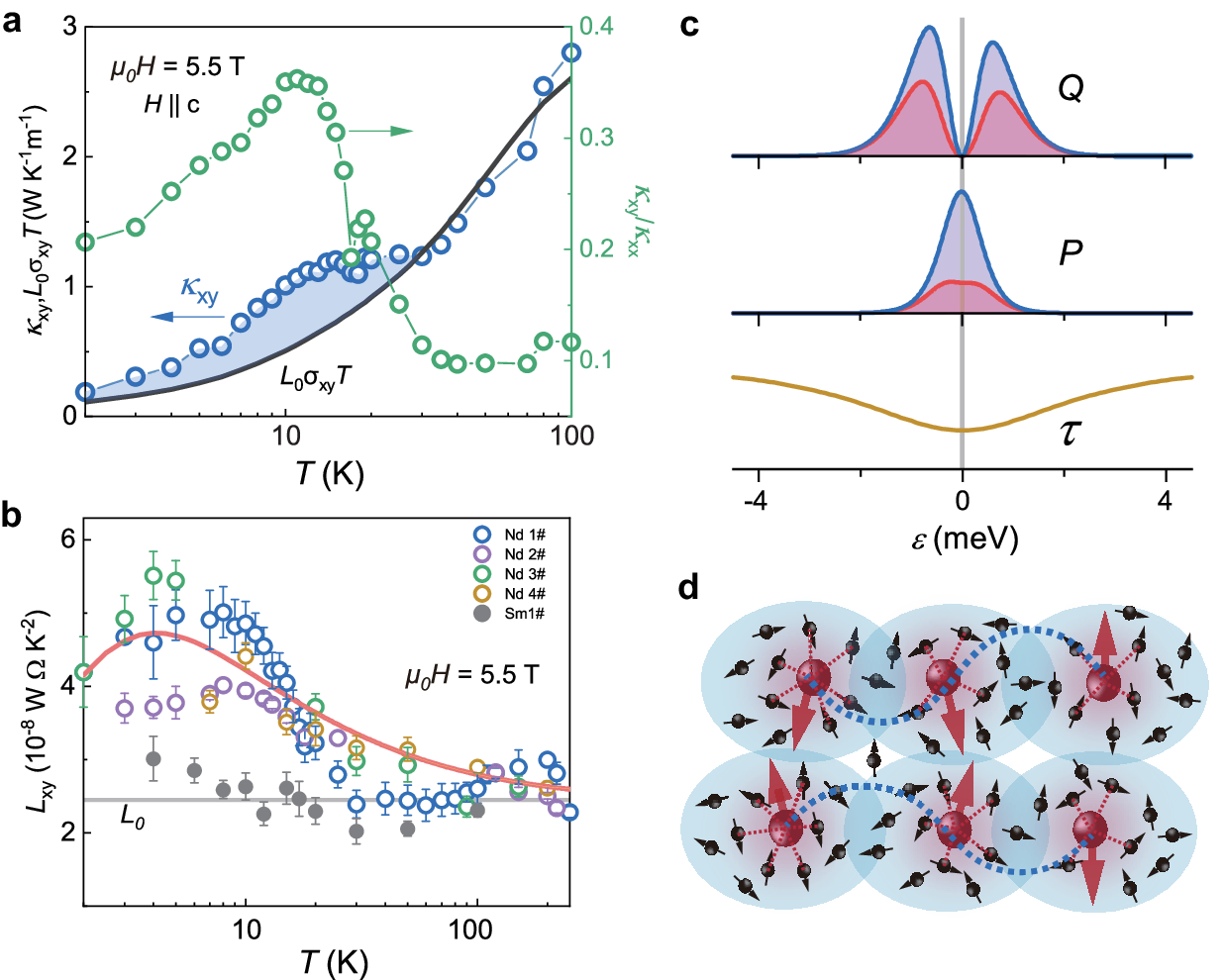}
		\caption{{\bf Finite $T$ deviation from Wiedemann-Franz law and the Kondo-type scattering.} {\bf a,} $T$-dependence of the in-plane thermal Hall $\kappa_{xy}$ (blue circles), the thermal Hall angle $\kappa_{xy}$/$\kappa_{xx}$ (green circles) and $L_0\sigma_{xy}T$ (black curve). Data were measured on NdAlSi sample \#1 at $\mu_0H$ = 5.5\,T with $\nabla T \parallel a$ and $H \parallel c$. {\bf b,} $L_{xy}$ measured under $\mu_0H$ = 5.5\,T at varying temperatures. Data were collected from four NdAlSi crystals (\#1-\#4). Measurement results of the isostructural compound SmAlSi are also shown. The red curve denotes $L_{xy}(T)$ calculated based on our model of $\tau(\epsilon)$ (Supplementary Note 9). The error bars were determined from the uncertainties (i.e., noise level) in the measurements of $\Delta T_x$ and $\Delta T_y$ (details in Supplementary Note 1). {\bf c,} The energy distribution of relaxation time $\tau$ (bottom) with a minimum that is symmetrical with regard to the chemical potential; such feature of $\tau(\epsilon)$ stems from the high-order (Kondo-type) scattering between 4$f$ and 5$d$ electrons (Supplementary Note 9). Integrands of the transport coefficients $\kappa_{xy}$ ($Q$, top) and $\sigma_{xy}$ ($P$, middle) derived for $T$ = 3\,K (Supplementary Note 8); red and blue colors stand for the cases with and without taking into account the Kondo-type scattering, respectively. {\bf d,} Schematic of the coexistence of the RKKY interaction (blue shaded region) and the Kondo interaction (red shared region) in NdAlSi; the two compete and the former dominates. The red and black arrows represent the spins of Nd$^{3+}$ localized 4$f$ and itinerant 5$d$ electrons, respectively. Blue dashed lines denote the RKKY interaction between the local moments.}
		\label{fig:figure4}
	\end{center}
\end{figure}

\newpage
\clearpage
\renewcommand{\thefigure}{\textbf{\arabic{figure}}}
\renewcommand{\thetable}{\textbf{\arabic{table}}}
\setcounter{figure}{0}
\renewcommand{\figurename}{\textbf{Extended Data Fig.\,$\!\!$}}
\renewcommand{\tablename}{\textbf{Extended Data Table\,$\!\!$}}

\end{document}